\documentclass[10pt, twocolumn, pre, aps, superscriptaddress, showpacs]{revtex4-1}
\usepackage{amsmath, graphicx, subfigure, tikz}
\begin{document}

\title{Nematic Phase of the n-Component Cubic-Spin Spin Glass in d=3:\\
Liquid-Crystal Phase in a Dirty Magnet}

\author{E. Can Artun}
    \affiliation{T\"UBITAK Research Institute for Fundamental Sciences, Gebze, Kocaeli 41470, Turkey}
    \affiliation{Faculty of Engineering and Natural Sciences, Kadir Has University, Cibali, Istanbul 34083, Turkey}
\author{Deniz Sarman}
    \affiliation{Department of Physics, Bo\u{g}azi\c{c}i University, Bebek, Istanbul 34342, Turkey}
\author{A. Nihat Berker}
    \affiliation{Faculty of Engineering and Natural Sciences, Kadir Has University, Cibali, Istanbul 34083, Turkey}
    \affiliation{T\"UBITAK Research Institute for Fundamental Sciences, Gebze, Kocaeli 41470, Turkey}
    \affiliation{Department of Physics, Massachusetts Institute of Technology, Cambridge, Massachusetts 02139, USA}

\begin{abstract}
A nematic phase, previously seen in the $d=3$ classical Heisenberg spin-glass system, occurs in the n-component cubic-spin spin-glass system, between the low-temperature spin-glass phase and the high temperature disordered phase, for number of components $n\geq 3$, in spatial dimension $d=3$, thus constituting a liquid-crystal phase in a dirty (quenched-disordered) magnet.  This result is obtained from renormalization-group calculations that are exact on the hierarchical lattice and, equivalently, approximate on the cubic spatial lattice. The nematic phase completely intervenes between the spin-glass phase and the disordered phase.  The Lyapunov exponents of the spin-glass chaos are calculated from $n=1$ up to $n=12$ and show odd-even oscillations with respect to $n$.

\end{abstract}
\maketitle

\section{Cubic-Spin Spin-Glass System and Nematic Phase in a Dirty Magnet}

Spin-glass systems have an inherent quantifiable chaos under scale change \cite{McKayChaos,McKayChaos2,BerkerMcKay,McKayChaos4} and thus provide a universal classification and clustering scheme for complex phenomena \cite{classif}, as well as rich ordering phenomena such as spin-glass sponge ordering \cite{sponge}.  Spin-glass studies have been done overwhelmingly with Ising $s_i=\pm1$ spins.  However, a recent study \cite{Tunca} with classical Heisenberg spins $\vec s_i$ that can continuously point in $4\pi$ steradians found, instead of spin-glass order, nematic order, meaning a liquid-crystal phase in a dirty magnet.  In the currrent study, we continue in this direction, studying $n$-component cubic-spin spin-glasses.  Cubic spins can be considered as discretized and systematically extended versions of Heisenberg spins.

\begin{figure}[ht!]
\centering
\includegraphics[scale=0.55]{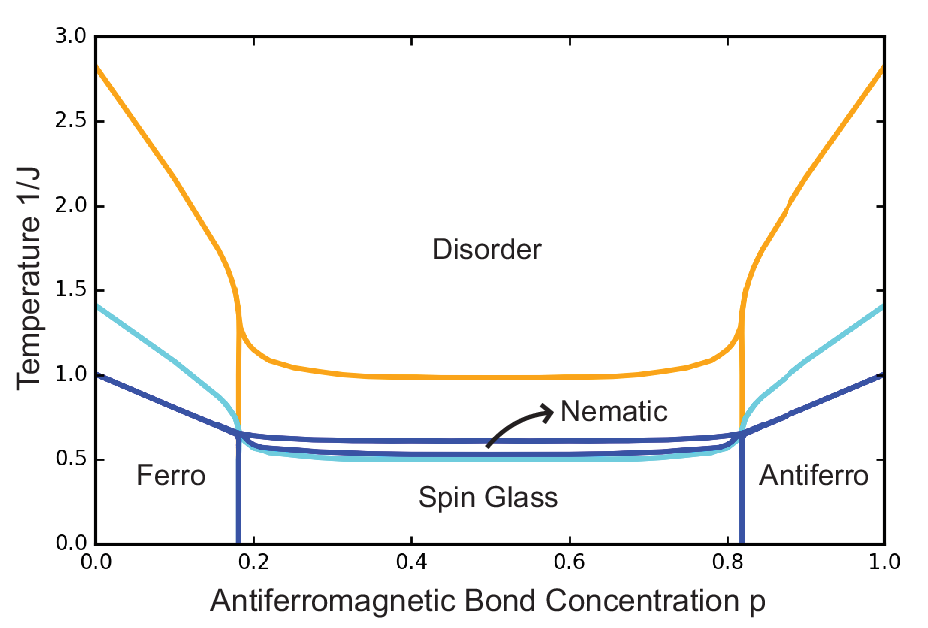}
\caption{Calculated phase diagrams for the cubic-spin spin-glass systems in spatial dimension $d=3$.  The phase diagrams are, from top to bottom, for number of components $n=1,2,3$ components, meaning $2n$ states.  A nematic phase appears for $n=3$, persisting for $n\geq3$, as seen in Fig.2.}
\end{figure}

\begin{figure}[ht!]
\centering
\includegraphics[scale=0.55]{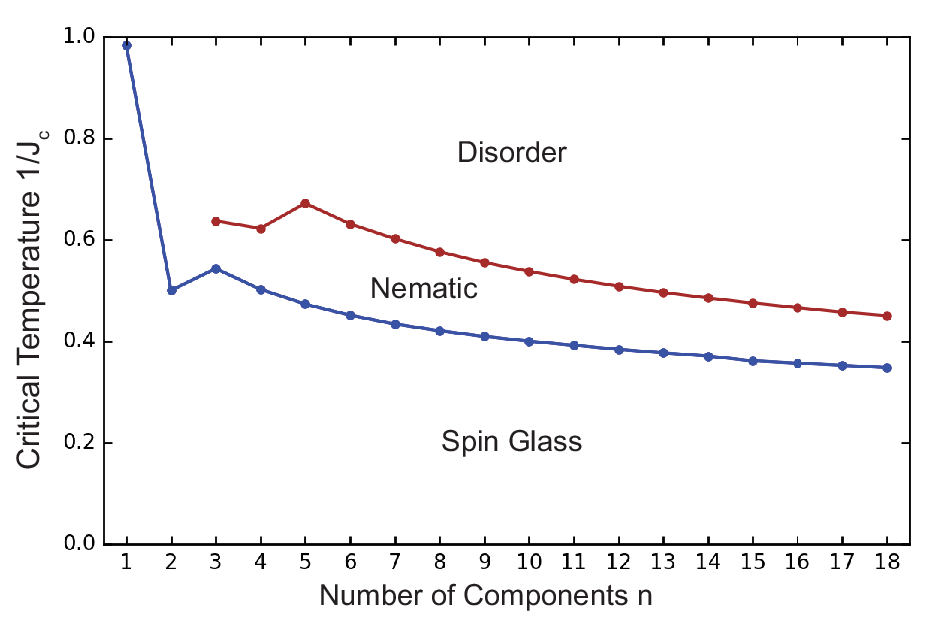}
\caption{Nematic and spin-glass transition temperatures at the ferro-antiferro symmetric value of $p=0.5$, as a function of the number $n$ of spin components.  It is seen that the nematic phase persists in $n\geq3$.}
\end{figure}

\begin{figure}[ht!]
\centering
\includegraphics[scale=0.5]{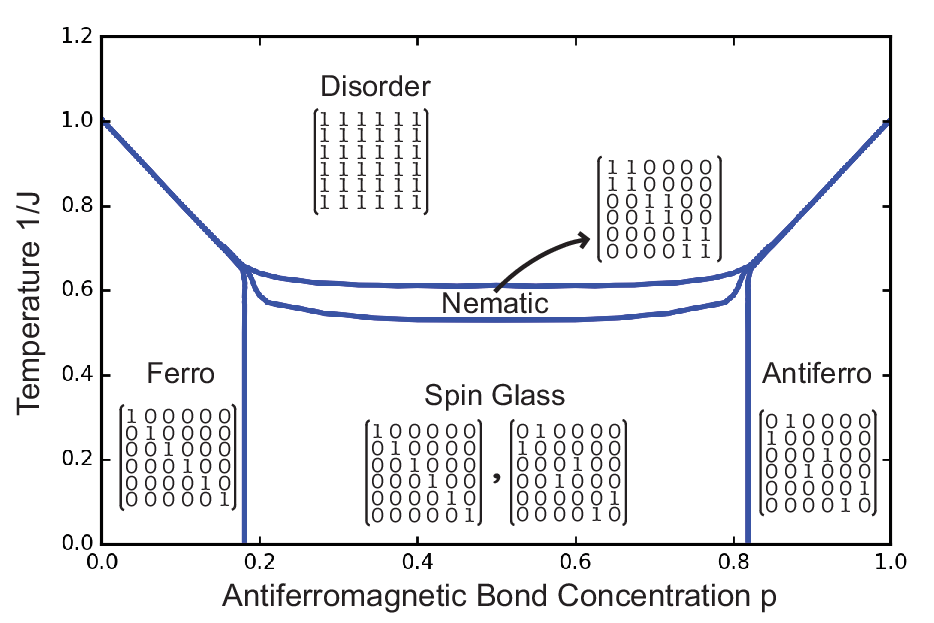}
\caption{Renormalization-group sinks of the phases for $n=3$.  The sink fixed-point values of the nearest-neighbor Hamiltonian exponentiated, namely the transfer matrix, are shown. All points in a given phase flow, under renormalization-group, to its corresponding sink, which epitomizes the ordering of the phase.  Except for the spin-glass phase, all phases flow to their respective single transfer matrix, shown in this figure.  The spin-glass phase flows to a fixed distribution of diverging ferromagnetic and antiferromagnetic interactions, shown in Fig. 4.  This fixed-point structure in this Fig. 3, underpinning the phase diagrams, occurs for all $n\geq 3$.}
\end{figure}

\begin{figure}[ht!]
\centering
\includegraphics[scale=0.5]{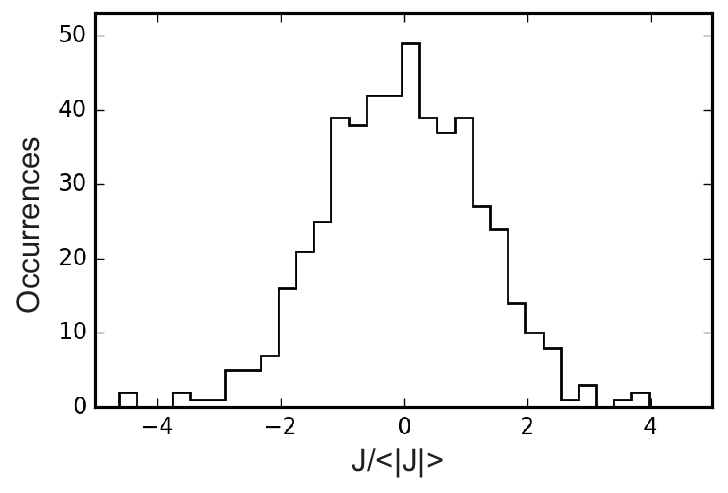}
\caption{The fixed distribution of the sink of the spin-glass phase.  The interaction $J$ is extracted from the $2 \times 2$ transfer matrix of the spin component $\hat{u}$ that shows spin-glass order.}
\end{figure}

\begin{figure*}[ht!]
\centering
\includegraphics[scale=0.5]{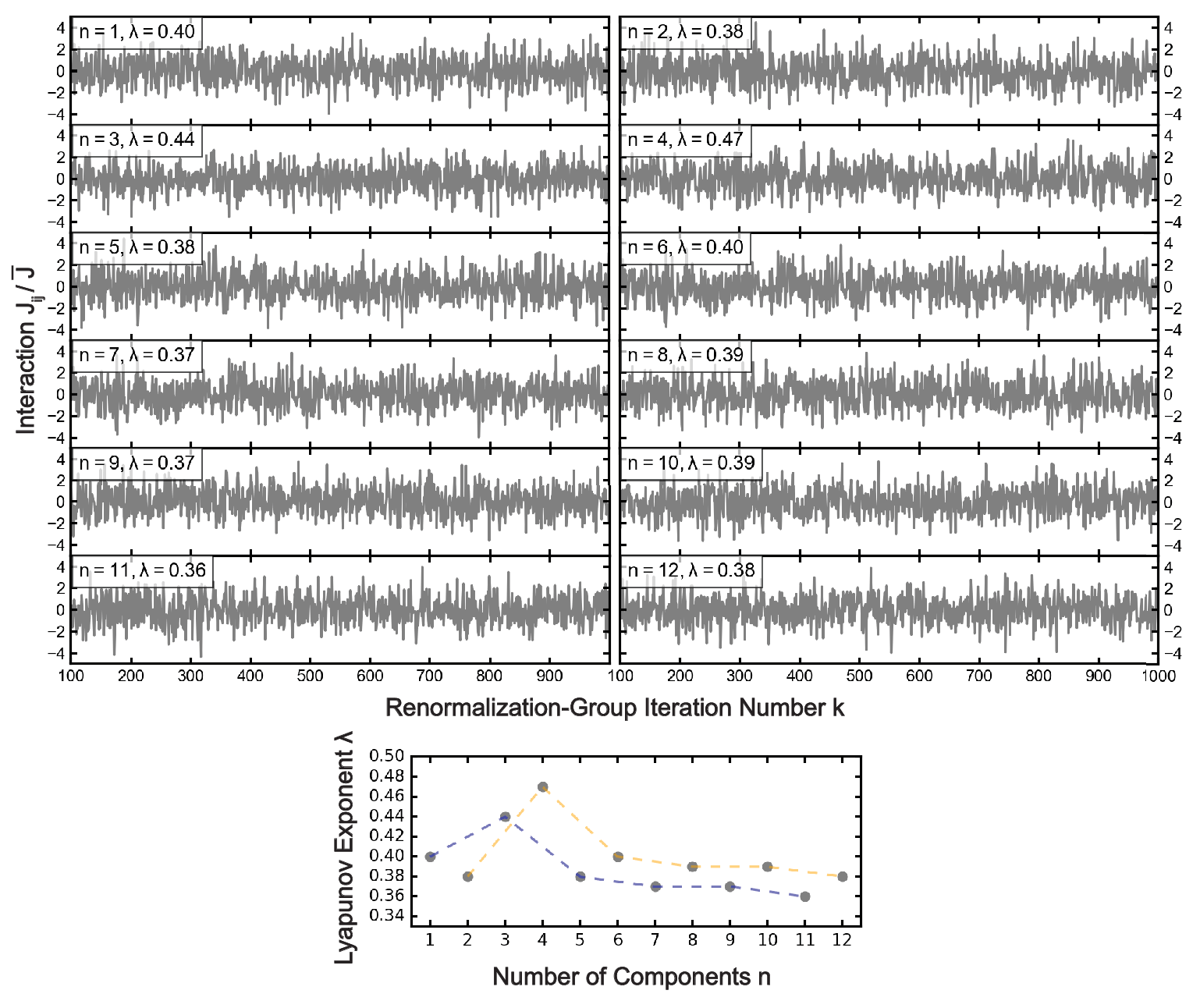}
\caption{Chaotic renormalization-group trajectories of the spin-glass phases and their Lyapunov exponents.  The Lyapunov exponents show odd-even oscillations with respect to the number of cubic components $n$.}
\end{figure*}

The n-component cubic-spin spin-glass system is defined by the usual Hamiltonian, where $\beta=1/kT$,
\begin{equation}
-\beta \mathcal{H}=\sum_{\langle ij \rangle} J_{ij} \vec s_i \cdot \vec s_j \equiv \sum_{\langle ij \rangle} -\beta {\cal H}_{ij} \,,
\end{equation}
but where $\vec s_i$ can be in $2n$ different states $\pm \hat{u}$ at each site $i$, $\hat{u}$ being a unit Cartesian vector.  The sum is over nearest-neighbor pairs of site $\langle ij \rangle$. The interaction $J_{ij}$ is ferromagnetic $+J>0$ or antiferromagnetic $-J$ with probabilities $1-p$ and $p$, respectively.

The cubic-spin spin-glass system studied in this work is of particular interest, because it is generalizes the much studied single-component simple spin Ising model to a realistic vector-spin system with $n$ vector components.  It is a rich system, since each of the $n$-component systems has phase transition behavior in a different universality class.  As seen below, it is readily treatable by renormalization-group theory, yielding the complex novel result of a band of liquid-crystal phase above the spin-glass phase.  This means a liquid-crystal phase occurring in a dirty magnet.

\section{Exactly Solved Hierarchical Model and Equivalent Migdal-Kadanoff Procedure}

The Migdal-Kadanoff approximation \cite{Migdal,Kadanoff} and its algebraically equivalent, exactly solved hierarchical model \cite{BerkerOstlund,Kaufman1,Kaufman2,BerkerMcKay} has been extensively described elsewhere.  Operationally, this renormalization-group transformation consists in decimating inside $b$ bonds in series and then adding $b^{d-1}$ decimated bonds in parallel, where $b$ is the length-rescaling factor ($b=3$ in this work). It is a very simply done, very effective, and most extensively used \cite{Clark,Kotorowicz,ZhangQiao,Jiang,Chio,Myshlyavtsev,Derevyagin,Shrock,Monthus,Sariyer,classif} renormalization-group transformation.  For quenched-random systems, as in the present work, starting with a large distribution (of 10,000 realizations in this work), a renormalized distribution is generated by randomly associating $b^d$ nearest-neighbor terms, 10,000 times.  This is the solution of an actual physical realization.\cite{Kraichnan,Flory,Kaufman,Lloyd}  Following the renormalization-group trajectories of the distributions, the phases and phase boundaries are obtained, as shown below. Thus, our analysis is done by analyzing the renormalization-group trajectories of 20,000 interactions, which is a gigantific calculation.

The transformation is best implemented algebraically by writing the exponentiated nearest-neighbor Hamiltonian, namely the transfer matrix between two neighboring sites, namely
\begin{multline}
\textbf{T}_{ij} \equiv e^{-\beta {\cal H}_{ij}} = e^{J \vec s_i \cdot \vec s_j} = \\
\left(
\begin{array}{cccccc}
e^J & e^{-J} & 1 & 1 & 1 & 1\\
e^{-J} & e^J & 1 & 1 & 1 & 1\\
1 & 1 & e^J & e^{-J}& 1 & 1 \\
1 & 1 & e^{-J} & e^J & 1 & 1 \\
1 & 1 & 1 & 1 & e^J & e^{-J} \\
1 & 1 & 1 & 1 & e^{-J} & e^J \end{array} \right),
\end{multline}
where the consecutive states are plus and minus directions of the sequence of cartesian unit vectors $\hat{u}$.  The transfer matrix is given here for $n=3$ components.  The generalization to arbitrary $n$ is obvious, and used in this work.  The decimation step consists in matrix-multiplying $b$ transfer matrices.  The bond-moving step consists in taking, after decimation, the power of $b^{d-1}$ of each element in the transfer matrix.  Here $b$ is the length-rescaling factor of the renormalization-group transformation.

\section{Emergent Nematic Phase in Calculated Phase Diagrams}

Calculated phase diagrams are shown in Fig. 1, for the cubic-spin spin-glass systems in spatial dimension $d=3$.  The phase diagrams are for number of components $n=1,2,3$, meaning $2n$ states.  A nematic phase appears for $n=3$, persisting for $n\geq3$, as seen in Fig.2. The renormalization-group sinks of the phases are shown in Fig. 3 for $n=3$.  The sink fixed-point values of the nearest-neighbor Hamiltonian exponentiated, namely the transfer matrix, Eq. (2), are shown. All points in a given phase flow, under renormalization-group, to its corresponding sink, which epitomizes the ordering of the phase.  Except for the spin-glass phase, all phases flow to their respective single transfer matrix, shown in this figure.  The spin-glass phase flows to a fixed distribution of diverging ferromagnetic and antiferromagnetic interactions, shown in Fig. 4.

The renormalization-group trajectories starting in a given phase unmistakably reach their respective sinks within 8 renormalization-group iterations, including the case of the nematic phase.  Renormalization-group trajectories starting close to a phase boundary spend iterations close to the unstable fixed point controlling the phase boundary, but reach  the sink within 15 iterations for the accuracy of our calculation.  The calculational uncertainy in our work is smaller than the thickness of all of the phase boundary lines in our figures.

The renormalization-group trajectories in the spin-glass phase are chaotic, as shown in Fig. 5.  The strength of chaos under scale change \cite{McKayChaos,McKayChaos2,BerkerMcKay,McKayChaos4} is measured by the Lyapunov exponent \cite{Collet,Hilborn},
\begin{equation}
\lambda = \lim _{n\rightarrow\infty} \frac{1}{n} \sum_{k=0}^{n-1} \ln \Big|\frac {dx_{k+1}}{dx_k}\Big|\,,
\end{equation}
where $x_k = J_{ij}/\overline{J}$ at step $k$ of the renormalization-group trajectory and $\overline{J}$ is the average of the absolute value of the interactions in the quenched random distribution.  We calculate the Lyapunov exponents by discarding the first 100 renormalization-group steps (to eliminate crossover from initial conditions to asymptotic behavior) and then using the next 900 steps, shown in Fig. 5.  The initial $(J)$ values do not matter, as long as they are within the specified spin-glass phase. As seen in Fig. 5, the calculated Lyapunov exponents show odd-even oscillations with respect to the number of cubic components $n$.  Such oscillations have been seen previously, for example, as a function of spin $s$ value, in critical temperatures of the Ising magnet.\cite{Berker}

In addition to chaos, the renormalization-group trajectories show asymptotic strong-coupling behavior \cite{Demirtas},
\begin{equation}
\overline{J'} = b^{y_R}\, \overline{J}\,,
\end{equation}
where $y_R >0$ is the strong-coupling runaway exponent \cite{Demirtas}.  Again using 900 renormalization-group steps after discarding 100 steps, we calculated the runaway exponent and find the same value of $y_R = 0.24$ for all $n$.  The fact that $y_R$ is much less than $d-1=2$ shows that this spin-glass order in very unsaturated order. In the compact (saturating) ordering of systems without quenched randomness, $y_R = d-1$ reflects the energy of the smooth interface between oppositely aligned renormalization-group cells.\cite{BerkerW}

The sink values of the nematic phase shows that the system aligns along one of the $\pm \hat{u}$, irrespective of the plus or minus direction, which is a nematic phase.  This is a spin-space discretized version of the nematic phase in the classical Heisenberg spin-glass system, where the nematic alignment is along one of the continuum of $4\pi$ steradians, again irrespective of $\pm$ direction.  The nematic phase intercedes, along the entire length, between the low-temperature spin-glass phase and the high-temperature disordered phase.  This renormalization-group and phase diagram picture, detailed here for $n=3$, occurs for all $n\geq 3$.

\section{Conclusion}

We have calculated the phase diagram of the cubic-spin spin-glass system in $d=3$.  We find that a nematic phase, such as seen in the classical Heisenberg model with continuously orientable spins, emerges from the chaotic spin-glass phase, for number of component $n\geq 3$.  The calculated Lyapunov exponents $\lambda$, measuring the strength of the spin-glass chaos, show odd-even oscillations with respect to the number of cubic spin components $n$, whereas the strong-coupling runaway exponent $y_R$ has the same low value of $0.24\ll d-1=2$ for all $n$, showing the much unsaturated order of the spin-glass phase.

\begin{acknowledgments}
Support by the Kadir Has University Doctoral Studies Scholarship Fund and by the Academy of Sciences of Turkey (T\"UBA) is gratefully acknowledged.
\end{acknowledgments}

\end{document}